\documentclass[twocolumn]{revtex4}
\usepackage{graphicx,epsf}
\usepackage{color}
\usepackage{bm}
\usepackage{dcolumn}
\usepackage{amsfonts}
\usepackage{mathrsfs}

\usepackage{subfig}

\begin{document}

\title{On Nonlinear Scattering of Drift Wave by Toroidal Alfv\'en Eigenmode in Tokamak Plasmas}

\author{ Liu Chen$^{1,2,3}$, Zhiyong Qiu$^{1,3}$, and Fulvio Zonca$^{3, 1}$}

\affiliation{$^1$Institute for    Fusion Theory and Simulation, School of Physics, Zhejiang University, Hangzhou, P.R.C\\
$^2$Department of   Physics and Astronomy,  University of California, Irvine CA 92697-4575, U.S.A.\\
$^3$ Center for Nonlinear Plasma Science and   C.R. ENEA Frascati, C.P. 65, 00044 Frascati, Italy}

\begin{abstract}

Using electron drift wave (eDW) as a paradigm model, we have investigated analytically direct wave-wave interactions between a test DW and ambient toroidal Alfv\'en eigenmodes (TAE) in toroidal plasmas,  and their effects  on the stability of the   eDW. The nonlinear effects enter via scatterings to short-wavelength electron Landau damped kinetic Alfv\'en waves (KAWs). Specifically, it is found that  scatterings to upper-sideband KAW lead to stimulated absorption of eDW. Scatterings to the lower-sideband KAW, on the contrary,  lead to its spontaneous emission. As a consequence, for typical parameters and fluctuation intensity, nonlinear scatterings by TAE have negligible net effects on   the eDW stability;  in contrast to  the ``reverse"  process investigated in    Ref. [Nuclear Fusion {\bf 62}, 094001 (2022)], where it is shown that  nonlinear scattering by ambient eDW may lead to  significant damping of   TAE. 

\end{abstract}


\maketitle

\section{Introduction}
\label{sec:intro}

 Drift wave (DW) \cite{WHortonRMP1999} and shear Alfv\'en wave (SAW) \cite{CZChengAP1985,KWongPRL1991,LChenRMP2016}   are two fundamental electromagnetic oscillations in magnetized plasmas such as tokamaks. DWs are, typically, electrostatic fluctuations excited by thermal plasma density and/or temperature nonuniformities. Consequently, DWs have frequencies, perpendicular wavelengths and parallel wavelengths comparable, respectively, to the thermal plasma diamagnetic drift frequencies, thermal ion Larmor radii and the system size.  SAWs, meanwhile, are electromagnetic fluctuations and, typically, manifest themselves as Alfv\'en eigenmodes (AEs) located  within the frequency gaps of SAW continuous  spectra \cite{CZChengAP1985}.   For typical tokamak parameters, AE frequencies could be an order of magnitude higher than those of DWs, and, thus, spontaneous excitations of AEs often involve resonances with superthermal energetic particles (EPs); e.g., alphas in a D-T fusion plasma. AEs, thus, have perpendicular wavelengths in the order of EP Larmor radii and parallel wavelengths in the order of system size. In short, we may describe DWs as low-frequency micro-scale fluctuations; while AEs  are meso-scale fluctuations at higher frequencies but still much lower than the ion cyclotron frequency.  Since both DWs and AEs are intrinsic fluctuations in magnetic confined fusion plasmas and have routinely been observed in tokamak plasmas, it is, thus, natural to inquire whether and how these two kinds of fluctuations may interact and what the potential  implications of these cross-scale interactions could be.  Recently, we have investigated such interactions via the channel of nonlinear wave scatterings between toroidal Alfv\'en eigenmode (TAE) \cite{CZChengAP1985} and, as a paradigm model, electron drift wave (eDW).  Interactions of DW turbulence and AEs have attracted significant interest in the recent years due to observed stabilization of tokamak turbulence by fast ions \cite{JCitrinPPCF2023,SMazziNP2022}. However, some fundamental aspects remain to be clarified and understood concerning the underlying physics processes.
 
 One important aspect  when extrapolating from present day devices to reactor relevant fusion plasmas is the EP characteristic energy and normalized orbit width, which are responsible of remarkably different EP dynamic responses in the two cases \cite{LChenRMP2016}. Another aspect concerns whether the predominant cross-scale coupling process is direct or indirect. In the first group is either stimulated or spontaneous wave-wave coupling.  Of the second type are processes mediated by zonal structures \cite{FZoncaPPCF2015}, e.g., zonal flows and fields \cite{LChenPRL2012,PDiamondPPCF2005}, including phase space zonal structures \cite{FZoncaNJP2015,MFalessiPoP2019}.
 
An example of direct coupling is the TAE/ITG (ion temperature gradient) induced scattering, where EP may excite TAE by inverse ion Landau damping in the presence of finite amplitude ITG turbulence \cite{VMarchenkoPoP2022}. This mechanism has been invoked to explain the observed excitation of marginally stable TAE in gyrokinetic simulations of ITG \cite{ADiSienaNF2019},  which then enhance the level of zonal flows and eventually yield to an appreciable reduction of ITG induced turbulence transport \cite{JCitrinPRL2013}.  In this work, we further explore the DW-AE direct coupling channel via nonlinear wave scatterings using the  eDW paradigm \cite{LChenNF2022} with the aim of developing a comprehensive gyrokinetic description of theses precesses and of gaining insights into their possible impact on turbulent transport.

There are two types of  direct  nonlinear interactions between TAE and eDW. The first type involves the scattering of a test TAE by ambient eDWs \cite{LChenNF2022}.  In this case, it was demonstrated that the TAE will suffer significant damping via nonlinearly generated upper and lower sidebands of short-wavelength electron Landau damped kinetic Alfv\'en waves (KAWs) \cite{AHasegawaPoF1976}.  This scattering process, thus, may be regarded as stimulated absorption. Furthermore, for typical parameters, it is found   that the nonlinear damping rate could be comparable to the growth rate of TAE instability excited by EPs.  The second type of nonlinear wave-wave interactions involve the scattering of a test eDW by ambient TAEs, and is the actual  focus of the present work. As will be shown in the following analysis, while the second type of scattering may be considered as the ``reverse" of the first one, the induced nonlinear damping/growth rate in this case is, in fact, negligible for typical parameters. Qualitatively speaking, while the nonlinearly  generated upper sideband KAW (UKAW) still gives rise to stimulated absorption, the nonlinearly generated lower sideband KAW (LKAW), however, gives rise to spontaneous emission (i.e., as in a parametric decay instability) \cite{MPollnauAPB2019}. Quantatively, these two effects tend to nearly cancel each other; leading to negligible net effect on the stability of eDW. 

The plan of this work is as follows. The theoretical model and governing equations are given in Sec. \ref{sec:model}. Section \ref{sec:sidebands} discusses the nonlinear generation of upper and lower KAW sidebands. Nonlinear dispersion relation of eDW in the presence of the finite-amplitude TAE is then derived and analyzed in Sec. \ref{sec:nl_dr}.  Section \ref{sec:conclusion} gives the final conclusions and discussions.

\section{Theoretical Model and Governing Equations}
\label{sec:model}

We consider a large-aspect-ratio and low-$\beta$ tokamak plasma with circular magnetic surfaces. Thus, $\epsilon\equiv r/R\ll1$ with $r$ and $R$ being,  respectively, the minor and major radii of the torus, and $\beta\sim O(\epsilon^2)\ll 1$ being the ratio between plasma and magnetic pressure.  We, furthermore, take the thermal background plasma to be Maxwellian, and adopt the   eDW paradigm model with finite density gradient but negligible temperature gradient  as well as trapped particle effects in order to simplify the theoretical analyses and, thereby, illuminate the underlying physics. 

The perturbed distribution function, $\delta f_j$ with $j=e, i$ 
\begin{eqnarray}
\delta f_j=-(e/T)_j F_{Mj} \delta\phi+\exp(-\bm{\rho}\cdot\nabla)\delta g_j,
\end{eqnarray}
obeys the nonlinear gyrokinetic equation \cite{EFriemanPoF1982}
\begin{eqnarray}
&&\left(\partial_t+v_{\parallel}\mathbf{b}\cdot\nabla+\mathbf{v}_d\cdot\nabla+\langle\delta \mathbf{u}_g\rangle_{\alpha}\cdot\right)\delta g_j\nonumber\\
&=&(e/T)_jF_{Mj}\left(\partial_t+i\omega_{*j}\right)\langle \exp(\bm{\rho}_j\cdot\nabla)\delta L\rangle_{\alpha}.\label{eq:gk}
\end{eqnarray}
 Here, $F_{Mj}$ is the Maxwellian distribution, $\bm{\rho}_j=\mathbf{b}\times\mathbf{v}/\Omega_j$, $\mathbf{b}\equiv \mathbf{B}_0/B_0$, $\Omega_j=(eB_0/mc)_j$, $\delta g_j$ is the non-adiabatic particle response,  $\mathbf{v}_d=\mathbf{b}\times[(v^2_{\perp}/2)\nabla\ln B_0+v^2_{\parallel}\mathbf{b}\cdot\nabla\mathbf{b}]$ is the magnetic drift velocity, $\langle\mathbf{A}\rangle_{\alpha}$ denotes the gyro-phase averaging of $\mathbf{A}$, $\omega_{*j}=-i(cT/eB_0)\mathbf{b}\times\nabla\ln N_j\cdot\nabla$ is the diamagnetic drift frequency due to the finite density gradient, 
\begin{eqnarray}
\langle \delta\mathbf{u}_j\rangle_{\alpha} = (c/B_0)\mathbf{b}\times \nabla \langle \exp(-\bm{\rho}_j\cdot\nabla)\delta L\rangle_{\alpha},
\end{eqnarray}
 and 
\begin{eqnarray}
\delta L=\delta\phi-v_{\parallel}\delta A_{\parallel}/c
\end{eqnarray}
  with $\delta\phi$ and $\delta A_{\parallel}$ being, respectively, the scalar and parallel component of the vector potential. Note that, with $\beta\ll1$, magnetic compression may be neglected; i.e., $\delta B_{\parallel}\simeq0$.
  
Meanwhile, the governing field equations are the quasi-neutrality condition
  \begin{eqnarray}
  \sum_{j=e,i} \left[ (N_0e^2/T)_j\delta\phi-e_j\left\langle (J_k\delta g)_j  \right\rangle_v      \right]=0,\label{eq:qn}
  \end{eqnarray}
 and the parallel Ampere's law $\nabla^2_{\perp}\delta A_{\parallel}=-(4\pi/c)\delta J_{\parallel}$.  Here, we note $J_k=J_0(k_{\perp}\rho)=\langle \exp(i\bm{\rho}\cdot\mathbf{k}_{\perp})\rangle_{\alpha}$ and $k^2_{\perp}=-\nabla^2_{\perp}$ should be understood as an operator. Furthermore, we note that, for SAW and KAW, instead of the Ampere's law, it is more convenient to use the following nonlinear gyrokinetic vorticity equation \cite{LChenJGR1991,LChenNF2001}
\begin{eqnarray}
&&i k_{\parallel} \delta J_{\parallel k} + (N_0e^2/T)_i \left(1-\Gamma_k\right)\left(\partial_t +i\omega_{*i}\right)_k\delta\phi_k\nonumber\\
&&  + i  \sum_j\left\langle e_j J_k \omega_d\delta g_j\right\rangle_v = \sum_{\mathbf{k}'+\mathbf{k}''=\mathbf{k}} \Lambda^{k'}_{k''} \left\{\delta A_{\parallel k'}\delta J_{k''}/c\right.\nonumber\\
&&\left. -e_j\left\langle \left(J_kJ_{k'}-J_{k''}\right)\delta L_{k'}\delta g_{k''j}\right\rangle_v\right\}.\label{eq:vorticity}
\end{eqnarray}
Here, $\Gamma_k\equiv I_0(b_k)\exp(-b_k)$, $b_k=k^2_{\perp}\rho^2_i$, $\rho^2_i=T_i/(m_i\Omega^2_i)$, $\omega_d=\bm{k}_{\perp}\cdot\bm{v}_d$ and $I_0$ is the modified Bessel function.  The first and second terms on the left hand side correspond, respectively, to the field line bending and inertia terms. Meanwhile, the third term corresponds to the curvature-pressure coupling term including the ballooning-interchange term and finite plasma compression. Note that, for TAE/KAW physics considered here, it can generally be ignored. The right hand side contains  the nonlinear terms,  where $\Lambda^{k'}_{k''}=(c/B_0) \mathbf{b}\cdot(\mathbf{k}''\times\mathbf{k}')$, and the first and second terms correspond, respectively, to the Maxwell and generalized gyrokinetic ion Reynolds stresses. 
Note that, since eDW is predominantly electrostatic, the Maxwell stress makes negligible contribution in the present analysis. 

\begin{center}
\begin{figure}
\includegraphics[scale=0.50]{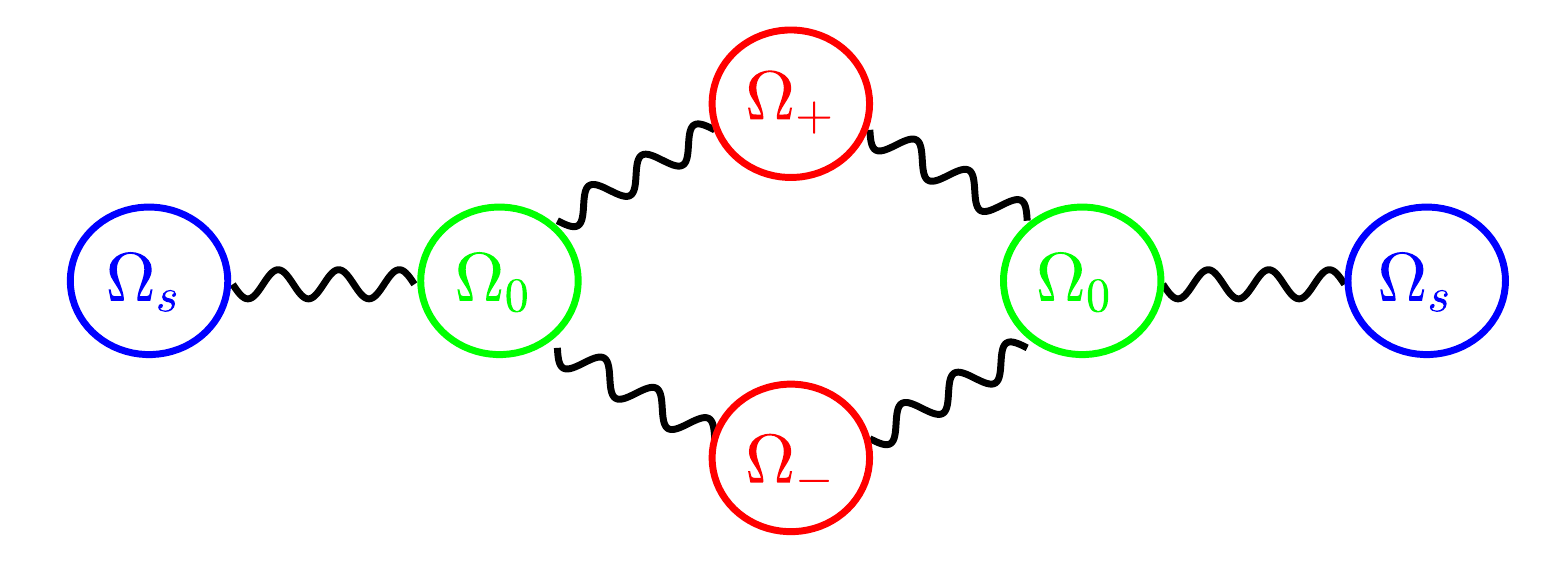} 
\caption{Schematic diagram of the two-step scattering processes analyzed in the present work. The test eDW, ambient TAE and nonlinearly generated KAW sidebands are  in blue, green and red, respectively.}\label{fig:schematic_diagram}
\end{figure}
\end{center}

We now consider the effects on eDW linear stability due to nonlinear scattering by TAE. Letting $\Omega_0(\omega_0,\mathbf{k}_0)$ and $\Omega_s=(\omega_s,\mathbf{k}_s)$ denote, respectively, a small but finite-amplitude TAE with toroidal mode number, $n_0$, and a test eDW with toroidal mode number, $n_s$. Thus, $|\omega_0|\simeq V_A/(2qR)$ with $V_A$ being the Alfv\'en speed and $q$ the safety factor, $\omega_s\sim\omega_{*e}$ the electron diamagnetic drift frequency, and  $|k_{s\theta}\rho_i|=|n_sq\rho_i/r|\sim O(1)$. Furthermore, we have, typically, $|\omega_s/\omega_0|<1$ and $|n_0/n_s|<1$.  That is, TAE and eDW are disparate both in spatial and temporal scales. Consequently, the sidebands nonlinearly generated by TAE and eDW;  i.e.,  $\Omega_{\pm}=(\omega_{\pm}, \mathbf{k}_{\pm})=\Omega_s\pm\Omega_0$, tend to have $|\omega_{\pm}|\simeq |\omega_0|$ and $|\mathbf{k}_{\pm}|\simeq |\mathbf{k}_s|$, and may be regarded as short-wavelength (high-$n$) KAWs.  $\Omega_{\pm}$, in turn, can interact with $\Omega_0$; resulting in the nonlinear modification of eDW dispersion relation and, thereby, of its stability properties.  The two-step scattering processes are illustrated schematically in Fig. \ref{fig:schematic_diagram}.
 The first-step scattering process, i.e., the nonlinear generation of KAW sidebands is analyzed in the following Sec. \ref{sec:sidebands}. Section \ref{sec:nl_dr} analyzes the second step scattering process and the resultant nonlinear eDW dispersion relation. 
  
\section{Nonlinear Generation of Upper and Lower Sidebands of Kinetic Alfv\'en Waves}
\label{sec:sidebands}

Let us first analyze the nonlinear generation of $\Omega_+$; i.e, UKAW. The analysis for LKAW is similar. For electrons, we let  $\delta g_{ke}=\delta g^{(1)}_{ke}+\delta g^{(2)}_{ke}$,  with superscripts ``(1)" and ``(2)" denoting, respectively,  the linear and nonlinear responses. Thus, from Eq. \ref{eq:gk}, we have
\begin{eqnarray}
\delta g^{(1)}_{ke}\simeq -\frac{e}{T_e} F_{Me}\left(1-\frac{\omega_{*e}}{\omega}\right)_k\delta\psi_k,\label{eq:l_e_saw}
\end{eqnarray}
where $\delta\psi_k=(\omega\delta A_{\parallel}/ck_{\parallel})_k$ is the effective potential due to the induced parallel electric field,  $-\partial_t \delta A_{\parallel}/c$,  and we have taken $|k_{\perp}\rho_e|\ll1$ and  the massless-electron $|\omega_k/k_{\parallel}v_{te}|\ll1$ limit, with $v_{tj}$ the thermal speed of the $j$-specie. In Eq. (\ref{eq:l_e_saw}), $k$ stands for the TAE/KAW modes;  viz., $\Omega_0$ and $\Omega_{\pm}$, and $\delta g^{(1)}_{se}\simeq 0$ as $\Omega_s$ is the predominantly electrostatic eDW mode. It then follows 
\begin{eqnarray}
\delta g^{(2)}_{+e}\simeq 0. \label{eq:nl_e_ukaw}
\end{eqnarray}
Meanwhile, for singly charged  ions  with $|\omega_k/k_{\parallel}v_{ti}|\gg1$ for all the modes considered here, TAE, KAW and eDW, we have 
\begin{eqnarray}
\delta g^{(1)}_{ki}\simeq \frac{e}{T_i}F_{Mi} J_k\delta\phi_k\left(1-\frac{\omega_{*i}}{\omega}\right)_k,\label{eq:linear_ion}
\end{eqnarray}
and
\begin{eqnarray}
\delta g^{(2)}_{+i}\simeq -i\frac{\Lambda^s_0}{2\omega_+} J_0J_s\frac{e}{T_i} F_{Mi}\left(\frac{\omega_{*i}}{\omega}\right)_s\delta\phi_s\delta\phi_0.\label{eq:nl_i_ukaw}
\end{eqnarray}
Substituting Eqs. (\ref{eq:l_e_saw}) to (\ref{eq:nl_i_ukaw}) into the quasi-neutrality condition, Eq. (\ref{eq:qn}),  it is possible to derive
\begin{eqnarray}
\delta\psi_+ = \sigma_{*+}\delta\phi_+ +i\frac{\Lambda^s_0}{2\omega_+} D_+\delta\phi_0\delta\phi_s, \label{eq:ukaw_qn}
\end{eqnarray}
where 
\begin{eqnarray}
\sigma_{*+}=\left[1+\tau-\tau \Gamma_+  (1-\omega_{*i}/\omega)_+\right]/(1-\omega_{*e}/\omega)_+,
\end{eqnarray}
and 
\begin{eqnarray}
D_+=\tau(\omega_{*i}/\omega)_s F_+/(1-\omega_{*e}/\omega)_+,
\end{eqnarray}
$\tau=T_e/T_i$, and $F_+=\langle J_0J_+J_sF_{Mi}\rangle_v/N_0$. Meanwhile, the nonlinear gyrokinetic vorticity equation, Eq. (\ref{eq:vorticity}),  yields
\begin{eqnarray}
&&\tau b_+\left[\left(1-\frac{\omega_{*i}}{\omega}\right)_+\frac{1-\Gamma_+}{b_+}\delta\phi_+ - \left(\frac{V^2_A k_{\parallel} b k_{\parallel}}{b \omega^2}\right)_+ \delta\psi_+     \right] \nonumber\\
&=& -i\frac{\Lambda^s_0}{2\omega_+} \gamma_+ \delta\phi_s\delta\phi_0,\label{eq:ukaw_vorticity}
\end{eqnarray}
where 
\begin{eqnarray}
\gamma_+=\tau\left[\Gamma_s-\Gamma_0 + (\omega_{*i}/\omega)_s (F_+-\Gamma_s)\right].
\end{eqnarray}
We note that,  $k_{\parallel}$  and  $b\propto k^2_{\perp}$  should be strictly considered as operators, and  are thus,   not commutative in, e.g., the field line bending term in Eq. (\ref{eq:ukaw_vorticity}).  Combining Eqs. (\ref{eq:ukaw_qn}) and (\ref{eq:ukaw_vorticity}) then yields the equation describing the nonlinear generation of $\Omega_+$ by $\Omega_0$ and $\Omega_s$; i.e., 
\begin{eqnarray}
\tau b_+ \epsilon_{A+} \delta\phi_+ = -i (\Lambda^s_0/2\omega_+)\beta_+ \delta\phi_s\delta\phi_0, \label{eq:ukaw_eq}
\end{eqnarray}
where
\begin{eqnarray}
\epsilon_{Ak}=\left(1-\frac{\omega_{*i}}{\omega}\right)_k \frac{1-\Gamma_k}{b_k} - \left(\frac{V^2_A}{b} \frac{k_{\parallel}b k_{\parallel}}{\omega^2}\right)_k\sigma_{*k} \label{eq:ukaw_operator}
\end{eqnarray}
is the linear SAW/KAW operator, and
\begin{eqnarray}
\beta_+&=&\tau(\Gamma_s-\Gamma_0) + \tau\left(\frac{\omega_{*i}}{\omega}\right)_s\nonumber\\
&\times&\left[F_+-\Gamma_s- \left(\frac{k_{\parallel}bk_{\parallel}}{\omega^2}\right)_+\frac{\tau V^2_A F_+}{(1-\omega_{*e}/\omega)_+}\right].
\end{eqnarray}

Nonlinear generation of $\Omega_-$ follows that of $\Omega_+$, and we, therefore, present only the main results. For electrons, we have, again, $\delta g^{(2)}_{-e}\simeq 0$, and, for ions, 
\begin{eqnarray}
\delta g^{(2)}_{-i}\simeq i\frac{\Lambda^s_0}{2\omega_-} J_0J_s\frac{e}{T_i} F_{Mi} \left(\frac{\omega_{*i}}{\omega}\right)_s\delta\phi_s\delta\phi^*_0.
\end{eqnarray}
The quasi-neutrality condition, Eq. (\ref{eq:qn}), yields,
 \begin{eqnarray}
 \delta\psi_-=\sigma_{*-}\delta\phi_--i(\Lambda^s_0/2\omega_-) D_- \delta\phi_s\delta\phi^*_0,\label{eq:qn_lkaw}
 \end{eqnarray}
with
\begin{eqnarray}
D_-=\tau(\omega_{*i}/\omega)_sF_-/(1-\omega_{*e}/\omega)_-,
\end{eqnarray}
and $F_-=\langle J_0J_-J_sF_{Mi}\rangle_v/N_0$.  Meanwhile, the nonlinear gyrokinetic vorticity equation, Eq. (\ref{eq:vorticity}), yields
\begin{eqnarray}
&&\tau b_- \left[ \left(1-\frac{\omega_{*i}}{\omega}\right)_-\frac{(1-\Gamma_-)}{b_-}\delta\phi_- - \left(\frac{V^2_A k_{\parallel}bk_{\parallel}}{b \omega^2}\right)_-\delta\psi_-    \right]\nonumber\\
&=& i\frac{\Lambda^s_0}{2\omega_-} \gamma_-\delta\phi_s\delta\phi^*_0,\label{eq:vorticity_lkaw}
\end{eqnarray}
and
\begin{eqnarray}
\gamma_-=\tau\left[\Gamma_s-\Gamma_0 + (\omega_{*i}/\omega)_s(F_- - \Gamma_s)\right].
\end{eqnarray}
Finally, from Eqs. (\ref{eq:qn_lkaw}) and (\ref{eq:vorticity_lkaw}), we have
\begin{eqnarray}
\tau b_-\epsilon_{A-}\delta\phi_- = i(\Lambda^s_0/2\omega_-)\beta_-\delta\phi_s\delta\phi^*_0,\label{eq:lkaw_eq}
\end{eqnarray}
and 
\begin{eqnarray}
\beta_-&=&\tau (\Gamma_s-\Gamma_0)+ \tau\left(\frac{\omega_{*i}}{\omega}\right)_s \nonumber\\
&\times&  \left[F_--\Gamma_s -\left(\frac{k_{\parallel} b k_{\parallel}}{\omega^2}\right)_-  \frac{\tau V^2_A F_-}{(1-\omega_{*e}/\omega)_-} \right].
\end{eqnarray}
We remark, again, that $\epsilon_{A\pm}$ in Eqs. (\ref{eq:ukaw_eq}) and (\ref{eq:lkaw_eq}) are KAW operators. That is, in terms of physics,  Eqs. (\ref{eq:ukaw_eq}) and (\ref{eq:lkaw_eq}) describe mode-converted KAWs ($\Omega_{\pm}$) driven by the nonlinear coupling between a TAE ($\Omega_0$) and eDW ($\Omega_s$).

\section{Nonlinear Dispersion Relation of Electron Drift Wave}
\label{sec:nl_dr}

We now analyze the second scattering process between $\Omega_{\pm}$ and $\Omega_0$ back into $\Omega_s$. Again, let us first consider the $\Omega_+$ channel; i.e., $\Omega_+  -  \Omega_{0}\rightarrow \Omega_s$. From the nonlinear gyrokinetic equation, Eq. (\ref{eq:gk}), we have, for electrons in the massless $|\omega_k/k_{\parallel}v_{te}|\ll 1$ limit and noting Eqs. (\ref{eq:l_e_saw}) and (\ref{eq:nl_e_ukaw}), 
\begin{eqnarray}
\delta g^{(2)}_{se,+}\simeq -i \frac{\Lambda^s_0}{2\omega_+}  \frac{e}{T_e}F_{Me} \delta\psi_+\delta\psi^*_0 \left[1+ \frac{k_{\parallel0}}{k_{\parallel s}} \frac{(\omega_{*e}-\omega)_s}{\omega_0} \right].
\end{eqnarray}
 Here, $\delta g^{(2)}_{se,+}$ denotes  nonlinear electron response of $\Omega_s$ due to $\Omega_+$ and $\Omega^*_0$ coupling.   For ions, meanwhile, we have
\begin{eqnarray}
\delta g^{(2)}_{si,+}\simeq i(\Lambda^s_0/2\omega_s) \left(J_+\delta\phi_+\delta g^{(1)*}_{0i} - J_0\delta\phi^*_0\delta g_{+i}\right).
\end{eqnarray}
Here, we note that $\delta g_{+i}=\delta g^{(1)}_{+i} + \delta g^{(2)}_{+i}$  is   given, respectively,  by Eqs. (\ref{eq:linear_ion}) and (\ref{eq:nl_i_ukaw}). $\delta g^{(2)}_{si,+}$ is then given by 
\begin{eqnarray}
\delta g^{(2)}_{si,+}& \simeq& \left[ i \frac{\Lambda^s_0}{2\omega_+}  J_0J_+\delta\phi_+\delta\phi^*_0-\frac{(\Lambda^s_0)^2}{4\omega_s\omega_+} J^2_0 J_s|\delta\phi_0|^2\delta\phi_s\right]    \nonumber\\
&\times&   \left(\frac{\omega_{*i}}{\omega}\right)_s \frac{e}{T_i}F_{Mi}. 
\end{eqnarray}
The analysis is similar for the $\Omega_-+\Omega_0\rightarrow \Omega_s$ scattering channel. Then, we have
\begin{eqnarray}
\delta g^{(2)}_{se,-}\simeq  i \frac{\Lambda^s_0}{2\omega_-}  \frac{e}{T_e}F_{Me} \delta\psi_-\delta\psi_0 \left[1+ \frac{k_{\parallel0}}{k_{\parallel s}} \frac{(\omega_{*e}-\omega)_s}{\omega_0} \right],
\end{eqnarray}
and 
\begin{eqnarray}
\delta g^{(2)}_{si,-}& \simeq&- \left[ i \frac{\Lambda^s_0}{2\omega_-}  J_0J_-\delta\phi_-\delta\phi_0+\frac{(\Lambda^s_0)^2}{4\omega_s\omega_-} J^2_0 J_s|\delta\phi_0|^2\delta\phi_s\right]    \nonumber\\
&\times&   \left(\frac{\omega_{*i}}{\omega}\right)_s \frac{e}{T_i}F_{Mi}. 
\end{eqnarray}

Substituting the $\delta g_{sj}=\delta g^{(1)}_{sj}+\delta g^{(2)}_{sj,+}+\delta g^{(2)}_{sj,-}$ for $j=e,i$ into the quasi-neutrality condition, Eq. (\ref{eq:qn}), of  the $\Omega_s$ mode, we then readily derive the following governing equation for $\delta\phi_s$;
\begin{eqnarray}
\epsilon_s\delta\phi_s &=& i(\Lambda^s_0/2\omega_+) \beta_{s+}\delta\phi^*_0\delta\phi_+ - i (\Lambda^s_0/2\omega_-) \beta_{s-}\delta\phi_0\delta\phi_-\nonumber\\
&&-\epsilon^{(2)}_s|\delta\phi_0|^2 \delta\phi_s.\label{eq:eDW_eq}
\end{eqnarray}
Here, $\epsilon_s$ is the eDW linear dielectric operator and, in the limit of adiabatic circulating electrons and neglecting trapped electrons,  is given by
\begin{eqnarray}
\epsilon_s=1+\tau-\tau\left\langle \left(\frac{\omega-\omega_{*i}}{\omega-k_{\parallel}v_{\parallel}-\omega_d}\right)_s \frac{F_{Mi}}{N_0} J^2_s\right\rangle_v;
\end{eqnarray}
and, in the lowest order, 
\begin{eqnarray}
\epsilon_s\simeq 1+\tau(1-\Gamma_s) + \tau \Gamma_s (\omega_{*i}/\omega)_s. 
\end{eqnarray}

Meanwhile, 
\begin{eqnarray}
\beta_{s\pm} = \tau\left(\frac{\omega_{*i}}{\omega}\right)_s F_{\pm} + \sigma_{*0}\sigma_{*\pm} \left[ 1+\frac{k_{\parallel 0}}{k_{\parallel\pm}} \frac{(\omega_{*e}-\omega)_s}{\omega_0}\right],
\end{eqnarray}
and
\begin{eqnarray}
\epsilon^{(2)}_s&=& \sum_{l=+,-} \left\{ \frac{F_2}{\omega_s\omega_l} +\sigma_{*0} \left[1+\frac{k_{\parallel 0}}{k_{\parallel l}} \frac{(\omega_{*e}-\omega)_s}{\omega_0}  \right] \right.\nonumber\\
&&\left.\times \left[\frac{F_l}{\omega^2_l (1-\omega_{*e}/\omega)_l} \right]\right\}\frac{\left(\Lambda^s_0\right)^2}{4} \tau\left(\frac{\omega_{*i}}{\omega}\right)_s.
\end{eqnarray}

Noting Eqs. (\ref{eq:ukaw_eq}) and (\ref{eq:lkaw_eq}) for, respectively, $\delta\phi_+$ and $\delta\phi_-$, Eq. (\ref{eq:eDW_eq}) can be formally expressed as
\begin{eqnarray}
\left(\epsilon_s+\epsilon^{(2)}_s |\delta\phi_0|^2\right)\delta\phi_s &=&\left[ \left(\frac{\Lambda^s_0}{2\omega_+}\right)^2 \frac{\beta^+_s\delta\phi^*_0\beta_+}{\tau b_+\epsilon_{A+}} \delta\phi_0 \right.\nonumber\\
 &&\hspace*{-3em}\left.+    \left(\frac{\Lambda^s_0}{2\omega_-}\right)^2 \frac{\beta^-_s\delta\phi_0\beta_-}{\tau b_-\epsilon_{A-}} \delta\phi^*_0  \right]\delta\phi_s;\label{eq:nl_eDW_eq}
\end{eqnarray}
which may be regarded as the nonlinear eigenmode equation of $\Omega_s$ (eDW) in the presence of finite-amplitude $\Omega_0$ (TAE) fluctuations. 

Equation (\ref{eq:nl_eDW_eq}), in general, needs to be solved numerically. We can, however, make analytical progress by employing the scale separation and obtain an analytical dispersion relation variationally. First, we adopt the ballooning-mode representation for $\delta\phi_s$;
\begin{eqnarray}
\delta\phi_s=\exp(in_s\xi) \sum_{m_s}\exp(-im_s\theta)\Phi_s(n_sq-m_s\equiv z_s),
\end{eqnarray}
where $\xi$ and $\theta$ are, respectively the toroidal and poloidal angles, and denote the spatial scales of TAE and eDW as, respectively, $\mathbf{x}_0$ and $\mathbf{x}_s$; such that $|\mathbf{x}_s|/|\mathbf{x}_0|\sim O(n_0/n_s)\ll1$. Multiplying Eq. (\ref{eq:nl_eDW_eq}) by $\delta\phi^*_s$ and integrating over $\mathbf{x}_s$, we can derive 
\begin{eqnarray}
D_s + \chi^{(2)}_s |\delta\phi_0(\mathbf{x}_0)|^2= R_++R_-, \label{eq:nl_eDW_eigen}
\end{eqnarray}
where 
\begin{eqnarray}
D_s=\left\langle \Phi^*_s(z_s)\epsilon_s\Phi_s\right\rangle_s
\end{eqnarray}
is the linear dielectric constant of $\Omega_s$, 
\begin{eqnarray}
\langle\Phi^*_s[A]\Phi_s\rangle_s&\equiv& \int^{1/2}_{-1/2} dz_s\sum_{m_s} \Phi^*_s [A]\Phi_s\nonumber\\
&=&\int^{\infty}_{-\infty} dz_s \Phi^*_s[A]\Phi_s
\end{eqnarray}
with the normalization $\langle|\Phi_s|^2\rangle_s=1$,
\begin{eqnarray}
\chi^{(2)}_s=\langle \Phi^*_s\epsilon^{(2)}_s\Phi_s\rangle_s, 
\end{eqnarray}
and 
\begin{eqnarray}
R_{\pm}&=&\left\langle \Phi^*_s\left(\frac{\Lambda^s_s}{2\omega_{\pm}}\right)^2 \beta^{\pm}_s\left\{\delta\phi^*_0\atop\delta\phi_0\right\} \frac{\beta_{\pm}}{(\tau b\epsilon_A)_{\pm}} \left\{\delta\phi_0\atop\delta\phi^*_0\right\} \Phi_s\right\rangle_s.\nonumber\\
&& \label{eq:R_pm}
\end{eqnarray}

Equation (\ref{eq:nl_eDW_eigen}) is formally the variational  nonlinear eDW dispersion relation in the presence of a finite-amplitude TAE given by $\delta\phi_0$. We will later analyze it further using a trial function for $\Phi_s(z_s)$. We now make same qualitative observations.  We note that $\chi^{(2)}_s$ is real and, in general, $R_{\pm}=Re(R_{\pm})+ i Im(R_{\pm})$. Thus, $\chi^{(2)}_s$ and $Re(R_{\pm})$ lead to nonlinear frequency shift; while $Im(R_{\pm})$ gives rise to nonlinear damping or growth. Focusing on $Im(R_{\pm})$ first, we observe, from Eq. (\ref{eq:R_pm}), that $Im(R_{\pm}) \propto Im(1/\epsilon_{A\pm})$; i.e., the imaginary component of the SAW/KAW operator, $\epsilon_{A\pm}$, given  by Eq. (\ref{eq:ukaw_operator}). Looking at Eq. (\ref{eq:ukaw_eq}) and letting 
\begin{eqnarray}
\delta\phi_+ &=& A_+(\mathbf{x}_0) \exp(in_s\xi)\nonumber\\
&\times& \sum_{m_s} \exp(-im_s\theta) \Phi_+(z_s\equiv n_s q-m_s),
\end{eqnarray}
we then have, recalling the scale separation between $\mathbf{x}_0$ and $\mathbf{x}_s$, 
\begin{eqnarray}
A_+(\mathbf{x}_0) \tau b_s\epsilon^s_{A+} \Phi_+(z_s) =-i\frac{\Lambda^s_0}{2\omega_+} \beta_+ \Phi_s(z_s)\delta\phi_0(\mathbf{x}_0).\label{eq:ukaw_average}
\end{eqnarray}
The same analysis can be carried out for $\delta\phi_-$ given by Eq. (\ref{eq:lkaw_eq}) step by step. Further simplification of Eq. (\ref{eq:ukaw_average}) and the analogue for the $\Omega_-$ sideband can be obtained noting that
\begin{eqnarray}
\epsilon^s_{A\pm} &=& \left(1-\frac{\omega_*}{\omega}\right)_{\pm}\frac{1-\Gamma_s}{b_s} - \left(\frac{V^2_A}{b_s} \frac{k_{\parallel s}b_sk_{\parallel s}}{\omega^2_{\pm}}\right) \sigma^s_{*\pm},\nonumber\\
\sigma^s_{*\pm} &\simeq& \left[1+\tau-\tau\Gamma_s\left(1-\omega_{*i}/\omega\right)_{\pm}\right]/(1-\omega_{*e}/\omega)_{\pm},
\end{eqnarray}
$b_s= b_{s\theta} (1+\hat{s}^2\partial^2_{z_s})$, $b_{s\theta}=k^2_{s\theta}\rho^2_i$, $\hat{s}=rq'/q$ denotes magnetic shear,  and $k_{\parallel s} = (n_sq-m_s)/(qR)=z_s/(qR)$. Since $|\hat{s}^2\partial^2_{z_s}|<1$ for moderately/strongly ballooning modes, $\epsilon^s_{A\pm}$ further reduces to 
\begin{eqnarray}
\epsilon^s_{A\pm}\simeq b_{s\theta}\frac{\partial \epsilon^s_{A\pm}}{\partial b_{s\theta}} \hat{s}^2\partial^2_{z_s} - \left(\frac{\omega_A}{\omega_{\pm}}\right)^2\sigma_{\pm s}(z^2_s-z^2_{\pm}). \label{eq:epsilon_s_A}
\end{eqnarray}
Here, $\omega_A=V_A/(qR)$, 
\begin{eqnarray}
\sigma_{\pm s}=\left[1+\tau-\tau\Gamma_s(b_{s\theta})(1-\omega_{*i}/\omega)_{\pm}\right]/(1-\omega_{*e}/\omega)_{\pm},
\end{eqnarray}
and 
\begin{eqnarray}
z^2_{\pm}=\left(\frac{\omega}{\omega_A}\right)^2_{\pm} \left(1-\frac{\omega_{*i}}{\omega}\right)_{\pm} \frac{1-\Gamma_s(b_{s\theta})}{b_{s\theta}}\frac{1}{\sigma_{\pm s}} <\frac{1}{4};
\end{eqnarray}
as $|\omega/\omega_A|^2_{\pm}\simeq 1/4$. Equation (\ref{eq:ukaw_average}),   along with $\epsilon^s_{A+}$ given by Eq. (\ref{eq:epsilon_s_A}),  indicates that the upper sideband is a   mode converted KAW at the high-$n$ Alfv\'en resonance layer $z_s=\pm z_+$. As noted in previous study of mode-converted KAW \cite{AHasegawaPoF1976}, for $\tau=T_e/T_i\sim 1$, the finite electron Landau damping as well as the Airy swelling of the amplitude dictate that the damping occur predominantly around $z=\pm z_+$.   Furthermore, the spectrum of eDW is typically broad, which implies that the spectrum of mode-converted KAW is correspondingly broad. Thus,  the energy absorption rate approximates that of the local Alfv\'en resonance via the causality constraint $Im(\omega_s, \omega_+, \omega_-)>0$; i.e., 
\begin{eqnarray}
Im\left(\frac{1}{\epsilon_{A+}}\right)\simeq -\pi\delta(\epsilon_{A+})\simeq -\pi \left(\frac{\omega_+}{\omega_A}\right)^2\frac{\delta (z^2_s-z^2_+)}{\sigma_{+s}}.
\end{eqnarray}
Similar processes occur for the $\Omega_-$ KAW; i.e., 
\begin{eqnarray}
Im\left(\frac{1}{\epsilon_{A-}}\right)\simeq  \pi\delta(\epsilon_{A-})\simeq  \pi \left(\frac{\omega_-}{\omega_A}\right)^2\frac{\delta (z^2_s-z^2_-)}{\sigma_{-s}}.
\end{eqnarray}
Consequently,
\begin{eqnarray}
Im(R_+)&=&-\left\langle \Phi^*_s\left(\frac{\Lambda^s_0}{2\omega_+}\right)^2\frac{\beta^+_s\delta\phi^*_0\beta_+}{\tau b_s}\right.\nonumber\\
&&\left.\times\left(\frac{\omega_+}{\omega_A}\right)^2\frac{\delta(z^2_s-z^2_+)}{\sigma_{+s}}\delta\phi_0\Phi_s\right\rangle,
\end{eqnarray}
and, omitted here, a similar corresponding expression can be obtained for $Im(R_-)$. 

To proceed further, we take the $|k_{0\perp}\rho_i|^2\ll 1$ limit but keep the finite $|\omega_s/\omega_0|<1$ correction. It is then straightforward to derive
\begin{eqnarray}
\beta^{\pm}_s\beta_{\pm}\delta(\epsilon_{A\pm})&\simeq& \tau(1-\Gamma_{s\theta}) \sigma_{s\theta}\frac{(\omega_{*e}-\omega)_s(\omega-\omega_{*i})_s}{\omega^2_0}\nonumber\\
&&\times  \frac{k_{\parallel\pm}}{k_{\parallel s}} \delta (\epsilon_{A\pm}). 
\end{eqnarray}
 Here, $\Gamma_{s\theta}=\Gamma_s(b_{s\theta})$ and $\sigma_{s\theta}=1+\tau(1-\Gamma_{s\theta})$.

\begin{center}
\begin{figure}
\includegraphics[scale=0.30]{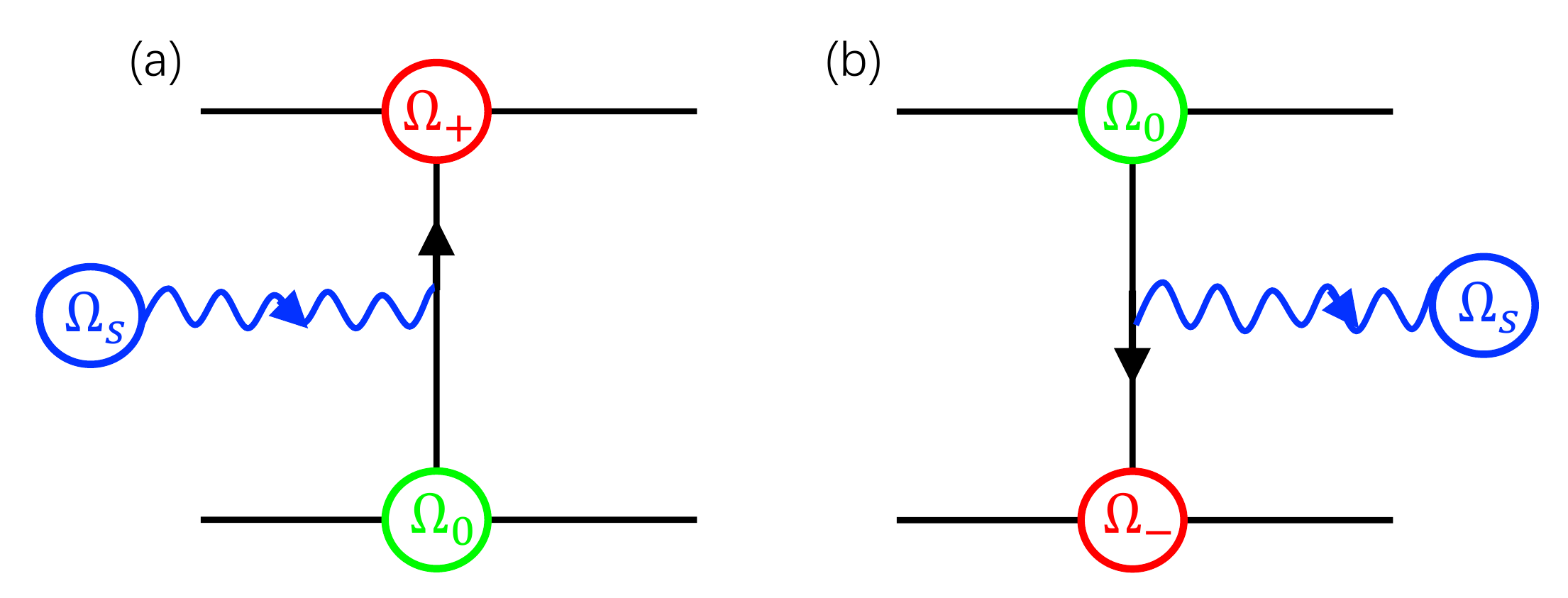} 
\caption{Sketch illustrating the nonlinear scattering processes of (a) stimulated absorption and (b) spontaneous emission.  Here, $\Omega_0$ is the finite-amplitude TAE, $\Omega_s$  is the test eDW, and $\Omega_{\pm}$ are, respectively, the upper and lower sideband KAWs. .}\label{fig:stimulated_absorption}
\end{figure}
\end{center}

The variational nonlinear   eDW dispersion relation, Eq. (\ref{eq:nl_eDW_eigen}),  then yields, letting $\omega_s=\omega_{sr}+i\gamma_s$ and $D_{sr}(\omega_{sr})=0$,
\begin{eqnarray}
&&\left(\gamma_s+\gamma^l_s\right)\frac{\partial}{\partial\omega_{sr}} D_{sr}=Im(R_++R_-)\nonumber\\
&\simeq&-\frac{\pi}{4\beta_i} \left(\frac{\Omega_i}{\omega_0}\right)^2\left|\frac{\delta B_{0\theta}}{B_0}\right|^2(1-\Gamma_{s\theta}) \sigma_{s\theta} \nonumber\\
&\times& \frac{(\omega_{*e}-\omega)_s(\omega-\omega_{*i})_s}{\omega^2_0} \left\langle \left[ \left(\frac{\omega_+}{\omega_A}\right)^2\frac{\delta(z^2_s-z^2_+)}{\sigma_{+s}}\right.\right.\nonumber\\
&&\left.\left. -   \left(\frac{\omega_-}{\omega_A}\right)^2\frac{\delta(z^2_s-z^2_-)}{\sigma_{-s}}\right] |\Phi_s|^2  \right\rangle_s.  \label{eq:nl_growth_rate}
\end{eqnarray}
Here, $\gamma^l_s$ is the linear damping/growth of eDW. Noting that $\partial D_{sr}/\partial\omega_{sr}>0$, $Im(R_+)<0$ and $Im(R_-)>0$, scatterings to UKAW and LKAW, thus, lead to, respectively, damping and growth of eDW. As illustrated in Fig. \ref{fig:stimulated_absorption}, one may qualitatively regard UKAW scattering as stimulated absorption, and LKAW scattering as spontaneous emission,  similar to the familiar parametric decay instability via a quasi-mode.   Since the rate of the two processes are nearly the same, the number of $\Omega_s$ quanta are approximately conserved, with a small overall effect on eDW growth/damping.    In Eq. (\ref{eq:nl_growth_rate}), we have also noted $ck_{0r}\delta\phi_0/B_0\simeq V_A\delta B_{0\theta}/B_0$.

To estimate the nonlinear damping/growth rate quantitatively, we adopt a trial function for $\Phi_s$ as $|\Phi_s|^2=(1/\sqrt{\pi}\Delta_s)\exp(-z^2_s/\Delta^2_s)$ with $\Delta_s>1$ for a typical moderately ballooning eDW. Equation (\ref{eq:nl_growth_rate}) then yields
\begin{eqnarray}
&&Im(R_++R_-) 
\simeq -\frac{\sqrt{\pi}}{4\beta_i} \left(\frac{\Omega_i}{\omega_0}\right)^2\left|\frac{\delta B_{0\theta}}{B_0}\right|^2\left(1-\Gamma_{s\theta}\right)\sigma_{s\theta} \nonumber\\
&\times&\frac{(\omega_{*e}-\omega)_s(\omega-\omega_{*i})_s}{\omega^2_0}\nonumber\\
&\times& \left[\left(\frac{\omega_+}{\omega_A}\right)^2\frac{1}{\sigma_{+s}z_+\Delta_s} -  \left(\frac{\omega_-}{\omega_A}\right)^2\frac{1}{\sigma_{-s}z_-\Delta_s}  \right]. 
\end{eqnarray}
Taking typical tokamak parameters, $\Omega_i/\omega_0\sim O(10^2)$, $\beta_i\sim O(10^{-2})$, $b_{s\theta}\sim  O(1)$, $|\omega_s/\omega_0|^2\sim O(10^{-1})$, and $|\Delta_sz_{\pm}| \sim O(1)$, we then find 
\begin{eqnarray}
|Im(R_++R_-)| <O(10^{5}) \left|\frac{\delta B_{0\theta}}{B_0}\right| ^2.
\end{eqnarray}
Noting that $\partial D_{sr}/\partial\omega_{sr}\sim 1/\omega_{sr}$ and $\gamma^l_{s}/\omega_{sr}\sim O(10^{-1})$ as, e.g., in the trapped electron mode \cite{GRewoldtCPC2007}, we then find that, for TAE fluctuations with $|\delta B_{0\theta}/B_0|^2\lesssim O(10^{-7})$ \cite{WHeidbrinkPRL2007}, the nonlinear contribution of damping/growth $\sim  |Im(R_++R_-)|\lesssim O(10^{-2})$, and should have negligible effects on the eDW stability.   We  also remark that one can, furthermore, straightforwardly show that the nonlinear frequency shift due to $\chi^{(2)}_s|\delta\phi_0|^2$ and $Re(R_++R_-)$ is also typically negligible.

\section{Conclusions and Discussions}
\label{sec:conclusion}

In this work, we have employed the nonlinear gyrokinetic equations and investigated analytically direct wave-wave interactions between a test electron drift wave (eDW) and ambient finite-amplitude toroidal Alfv\'en eigenmodes (TAEs) in low-$\beta$ circular tokamak plasmas.  Here, nonlinear scatterings generate upper and lower sidebands of mode-converted kinetic Alfv\'en waves (KAWs) at high todoial mode numbers which are typically damped by electrons around the mode conversion positions.  Furthermore, we find that  scattering  to upper-sideband KAW gives rise to stimulated absorption and, hence, damping of the eDW. Scattering  to lower-sideband KAW, on the other hand, gives rise to spontaneous emission and, thereby, growth of the eDW; i.e., TAE parametrically decays to eDW via the lower-sideband KAW quasi-mode. For typical tokamak parameters and TAE fluctuation intensity, our analysis indicates that the net effects on eDW stability properties should be negligible.  We remark again that, as noted in Sec. \ref{sec:intro}, the present results are different with  those obtained previously for the case of direct wave-wave interactions between a test TAE and ambient eDW \cite{LChenNF2022}.  In that case, both channels of scatterings to KAWs lead to stimulated absorption and, thereby, significant damping of the TAE.

As noted above, our analysis adopts the electron drift waves without temperature gradients as a paradigm model in order to simplify the analysis and delineate more clearly the underlying nonlinear physics mechanisms. It is clearly desirable to extend the investigations to include ion-temperature-gradient (ITG)  modes, trapped particle effects,  as well as other types of AEs; such as reversed shear Alfv\'en eigenmodes (RSAEs) \cite{HBerkPRL2001,FZoncaPoP2002} and beta-induced Alfv\'en eigenmodes (BAEs) \cite{WHeidbrinkPRL1993,FZoncaPPCF1996}.  While detailed analyses for such cases remain to be carried out, one may conjecture that the physical pictures outlined in the  current paradigm model should hold at least qualitatively. 

Finally, that the present results indicating negligible effects on eDW via direct wave-wave interactions with TAE suggests the possible significance of indirect   interaction via, e.g.,  the zonal structures consisting of flow, field and phase space nonlinearly generated by AEs \cite{FZoncaNJP2015,LChenPRL2012,DSpongPoP1994,ADiSienaJPP2021}. This interesting subject remains to be further investigated in the future.

\section*{Acknowledgement}
This work was  supported by  the National Science Foundation of China under Grant Nos. 12275236 and 12261131622, and ``Users of Excellence program of Hefei Science Center CAS" under Contract No. 2021HSC-UE016.
 This work was was supported by the EUROfusion Consortium, funded by the European Union via the Euratom Research and Training Programme (Grant Agreement No. 101052200 EUROfusion). The views and opinions expressed are, however, those of the author(s) only and do not necessarily reflect those of the European Union or the European Commission. Neither the European Union nor the European Commission can be held responsible for them.

\end{document}